\documentclass[prl,twocolumn,showpacs]{revtex4}

\usepackage{amsmath}
\usepackage[english]{babel}

\begin{document}

\title{Fault-tolerant holonomic quantum computation}
\author{Ognyan Oreshkov$^{(1)}$, Todd A. Brun$^{(1,2)}$, Daniel A. Lidar$%
^{(1,2,3)}$}
\affiliation{Departments of $^{(1)}$Physics, $^{(2)}$ Electrical Engineering, $^{(3)}$
Chemistry, Center for Quantum Information Science \& Technology, University
of Southern California, Los Angeles, California 90089}

\begin{abstract}
{We explain how to combine holonomic quantum computation (HQC) with
fault tolerant quantum error correction. This establishes the
scalability of HQC, putting it on equal
footing with other models of computation, while retaining the inherent
robustness the method derives from its geometric nature.}
\end{abstract}

\maketitle

\textit{Introduction}.---Holonomic quantum computation (HQC)
\cite{HQC} is an all-geometric method of computation that {makes use
of} non-Abelian generalizations of the Berry phase \cite{Berry84}.
{It is a fundamental alternative to other established models of
quantum computation (QC) \cite{QCmodels}, which in some sense lies
between the circuit and adiabatic models, in that it replaces the
dynamical gates of the circuit model by adiabatic holonomies. In HQC
states are encoded in the degenerate eigenspace of a Hamiltonian and
gates are realized by adiabatically varying the Hamiltonian along
suitable paths in parameter space, giving rise to geometric
transformations. HQC has attracted significant interest both because
of its {deep connections to gauge theory }\cite{WZ84} and its
potential practical advantages. It has been shown that due to its
geometric nature, the method is resilient to various types of errors
in the control parameters driving the evolution \cite{HQCrob} and
thus could provide a certain built-in robustness at the hardware
level. On the other hand, geometric phases are susceptible to
decoherence, a conclusion which also affects HQC
\cite{SarandyLidar:06}.

{No method of computation is scalable without the ability to
implement it fault-tolerantly, and HQC\ is no exception. However, unlike the
circuit and one-way models of quantum computation \cite{QCmodels} for which
fault tolerance proofs exist under reasonable noise models \cite{Got97,FT,AL06}, a
demonstration that HQC\ can be made fault tolerant under similar assumptions
has been lacking. Here we remedy this situation, establishing
the in-principle scalability of HQC.}

Fault-tolerant techniques guarantee {that, under appropriate
assumptions such as errors that are sufficiently uncorrelated and improbable}%
, an arbitrarily large computational task can be implemented with {%
relatively modest resource overhead that preserves the computational
complexity class}. This result, known as the threshold theorem, is
based on {the use of quantum error correcting codes} (QECCs)
\cite{QEC}---a {general} software solution to the {problems of noise
and decoherence} in quantum computers. In Ref.~\cite{WZL05}, {a
first step was taken to protect HQC against decoherence}, by
combining it with the method of decoherence-free subspaces (DFSs)
\cite{DFS}, leading to passive protection against certain types of
correlated errors. {However, this is not enough for fault tolerance,
since other error types could accumulate
detrimentally unless corrected}. Therefore, scalability of HQC {%
requires going beyond the scheme of Ref.~\cite{WZL05}, e.g., }by combining
the holonomic approach with \emph{active} error correction. Here we present
a scheme for fault-tolerant {HQC} using stabilizer {QECCs}
\cite{Stab}. {We begin by briefly reviewing} the basics of HQC and
the main ingredients of fault-tolerant QC. We then show how these
ingredients can be realized using holonomic transformations on {QECCs}%
, and prove that the construction is fault-tolerant. Finally, we discuss the
properties of the scheme and analyze its effect on the accuracy threshold.

\textit{Holonomic quantum computation}.---Let $\{H_{\lambda }\}$ be
an isodegenerate family of Hamiltonians on an $N$-dimensional
Hilbert space, which is continuously parameterized by a point
$\lambda $ in a differentiable control-parameter manifold
$\mathcal{M}$, i.e., $H_{\lambda
}=\sum_{n=1}^{R}\varepsilon _{n}(\lambda )\Pi _{n}(\lambda )$, where $%
\{\varepsilon _{n}(\lambda )\}_{n=1}^{R}$ are the $R$ different $d_{n}$-fold
degenerate eigenvalues of $H_{\lambda }$, ($\sum_{n=1}^{R}d_{n}=N$), and $%
\Pi _{n}(\lambda )$ are the projectors on the corresponding eigenspaces. If
the parameter $\lambda $ is changed adiabatically {(i.e.,
sufficiently slowly to prevent transitions between different eigenspaces),}
the unitary evolution that results from the action of the Hamiltonian $%
H(t):=H_{\lambda (t)}$ is
\begin{equation}
U(t)=\mathcal{T}\text{exp}(-i\int_{0}^{t}d\tau H(\tau ))=\oplus
_{n=1}^{R}e^{i\omega _{n}(t)}U_{A_{n}}^{\lambda }(t),
\label{adiabaticevolution}
\end{equation}%
where $\omega _{n}(t)=-\int_{0}^{t}d\tau \varepsilon _{n}(\lambda (\tau ))$
is a dynamical phase, and
\begin{equation}
U_{A_{n}}^{\lambda }(t)=\mathcal{P}\text{exp}(\int_{\lambda
(0)}^{\lambda (t)}\sum_{\mu }A_{n,\mu }d\lambda ^{\mu }).
\end{equation}%
Here $\mathcal{T}$ and $\mathcal{P}$ denote time- and path-ordering.
The adiabatic connection is $\sum_{\mu }A_{n,\mu }d\lambda ^{\mu }$,
where $\lambda ^{\mu }$ are local coordinates on $%
\mathcal{M}$ ($1\leq \mu \leq \text{dim}\mathcal{M}$), $A_{n,\mu }$
has matrix elements \cite{WZ84} $(A_{n,\mu })_{\alpha \beta
}=\langle n\alpha ;\lambda |\frac{\partial }{\partial \lambda ^{\mu }}%
|n\beta ;\lambda \rangle $, and $\{|n\alpha
;\lambda \rangle \}_{\alpha =1}^{d_{n}}$ is an orthonormal basis of the $n^{%
\text{th}}$ eigenspace of the Hamiltonian at the point $\lambda $.

When the path $\lambda (t)$ forms a loop $\gamma (t)$, $\gamma
(0)=\gamma (T)=\lambda _{0}$, the unitary matrix $U_{n}^{\gamma
}\equiv U_{A_{n}}^{\lambda }(T)=\mathcal{P}\text{exp}(\oint_{\gamma
}\sum_{\mu }A_{n,\mu }d\lambda ^{\mu })$ is called the holonomy
associated with the loop. When the $n^{\text{th}}$ energy level is
non-degenerate ($d_{n}=1$), the corresponding holonomy reduces to
the Berry phase \cite{Berry84}. {The space of all loops based on}
$\lambda _{0}$ is $L_{\lambda _{0}}(\mathcal{M})\equiv \{\gamma
:[0,T]\rightarrow \mathcal{M}|\gamma (0)=\gamma (T)=\lambda _{0}\}$.
The set $\text{Hol}(A_n)=\{U_n^{\gamma }|\gamma \in L_{\lambda
_{0}}(\mathcal{M})\}$ is a subgroup of $U(d_{n})$ called the
holonomy group. {If the dimension of }$\mathcal{M}${\ is
sufficiently large, non-Abelian holonomies can be used to implement
universal quantum computation over states encoded in one of the
degenerate eigenspaces of }$H(t)${\ \cite{HQC}.}

\textit{Fault-tolerant operations}.---We are concerned with standard
\cite{Stab} {and operator \cite{Bac06,OQEC}} stabilizer QECCs for
the correction of single-qubit errors, and the techniques for
fault-tolerant computation \cite{Got97,FT} on such codes. A quantum
information processing scheme is called fault-tolerant if a single
error occurring during the implementation of any operation
introduces at most one error per block of the code {(a block is a
set of qubits encoding one logical qubit). It is known \cite{Got97}
that fault-tolerant information processing is possible on any
stabilizer code and can be realized almost entirely in terms of
transversal operations---these are operations for which each qubit
in a block interacts only with the corresponding qubit from another
block or from a special ancillary state. In addition, it is required
that we are able to prepare and verify a special ancillary state.
Since single-qubit unitaries together with the \textquotedblleft
controlled-not\textquotedblright\ (CNOT) gate form a universal set
of gates, fault-tolerant computation can be realized entirely in
terms of single-qubit operations and CNOT operations between qubits
from different blocks, assuming that the special state can be
prepared reliably. Hence, our goal {is} to construct holonomic
realizations of these operations, as well as of the operations
needed for the preparation and use of the ancillary state.

The manner in which we combine HQC\ and QECCs is by embedding the
entire stabilizer code space or code subsystem into the ground space
of a two-level degenerate Hamiltonian $H_{\lambda }$ or in a
subsystem which is invariant under the action of $H_{\lambda }$. The
Hamiltonian is an element of either the stabilizer or the gauge
group of the code at the initial moment and remains an element of
the \textit{transformed} stabilizer or gauge group at every moment
during the computation. We perform computation in both the ground
and excited eigenspaces of $H_{\lambda }$. {Note that in this regard
we depart from the original HQC\ method \cite{HQC}, where
computation is performed entirely within a single eigenspace. }Even
though the geometric approach requires the use of degenerate
Hamiltonians which unavoidably couple qubits within the same block,
we show that propagation of single-qubit errors can be avoided.

\textit{The scheme}.---Consider a stabilizer code for the correction
of arbitrary single-qubit errors. We first show how to implement
encoded Clifford gates on such a code (Clifford gates are those that
preserve the Pauli group---the group of tensor products of Pauli
operators). It is known \cite{Got97} that these gates can be
realized using transversal Clifford operations. {For simplicity} we
restrict our attention to implementing transversal operations on the
first {qubit in each block}; the operations on the {remaining}
qubits can be obtained analogously, {and used to complete the
encoded Clifford gates on or between code blocks}. As a starting
Hamiltonian for implementing a single-qubit operation, we choose an
operator that is an element of the stabilizer, or the gauge group (for the case of subsystem codes \cite{OQEC}%
), and acts non-trivially on that qubit. Without loss of generality we can
write the initial Hamiltonian as $\widehat{H}(0)=-Z\otimes \widetilde{G}$,
where $X$, $Y$ and $Z$ are the Pauli matrices, and $\widetilde{G}$ is a tensor product of Pauli matrices on the rest of
the qubits in the block and possibly on qubits from other blocks if we are
in the middle of an entangling operation. (We can assume that $\widetilde{G}$
spreads {over} at most 4 blocks, since this is sufficient for
implementing transversally any encoded Clifford operation on stabilizer codes \cite%
{Got97}.) {Henceforth a} \textit{hat} denotes operators on all qubits
and \textit{tilde} on all qubits excluding the first one. {It is not
hard to show \cite{Ore08} that if the Hamiltonian is varied adiabatically so
that only the first factor changes,}
\begin{equation}
\widehat{H}(t)=-H(t)\otimes \widetilde{G},  \label{Ham1}
\end{equation}%
where $\text{Tr}\{H(t)\}=0$, {then the geometric transformation
resulting in each of the eigenspaces is equal to a local unitary operation on the first qubit}:
\begin{equation}
\widehat{U}(t)=U(t)\otimes \widetilde{I};\quad U(t)=U_{A_{0}}(t)\oplus
U_{A_{1}}(t),  \label{finalU}
\end{equation}%
where $U_{A_{n}}(t)=$$e^{\int_{0}^{t}d\tau \langle \phi _{n}(\tau )|\frac{d}{d\tau }%
|\phi _{n}(\tau )\rangle }|\phi _{n}(t)\rangle \langle \phi
_{n}(0)|$, with $|\phi _{n}(t)\rangle $ being the ground ($n=0$) and
excited ($n=1$) states of $H(t)$.

Equation (\ref{finalU}) is the basis of our construction. If $\widehat{H}(0)$
is minus an element of the stabilizer, then the code space belongs to its
ground space. Assuming that the encoded state has not undergone an error, by
varying the factor $H(t)$ adiabatically we can effectively generate any
single-qubit unitary transformation on the state (we show how to do this
below).
%We point out that when the final ground space is not the same as the
%initial one, the resulting transformation is an \textit{open-path} holonomy
%\cite{KAS06}.
If the initial Hamiltonian is an operator in the gauge group for the case of
subsystem codes, the non-erroneous state of the system can be a
superposition of ground and excited states. According to {Eq. (\ref%
{finalU})}, each of the ground and excited components will undergo the same
single-qubit unitary transformation $U(t)$, but in addition, a relative
phase of dynamical origin will accumulate between the two. This relative
phase is equivalent to an operation on the transformed gauge subsystem, and
therefore does not affect the encoded state.

\textit{Single-qubit operations}.---We now show how the {method above%
} can be used to generate a set of standard single-qubit gates. It turns out
\cite{Ore08} that any Hamiltonian of the type \eqref{Ham1} where
\begin{equation}
H(t)=f(t)Z+g(t)V_{\theta \pm }ZV_{\theta \pm }^{\dagger }\equiv
f(t)Z+g(t)H_{\theta \pm }  \label{interpolation}
\end{equation}%
with $f(T)=g(0)=0$, and $f(0),g(T)>0$, {gives rise in the adiabatic
limit to the geometric single-qubit unitary transformation}
\begin{equation}
V_{\theta \pm }=\frac{1}{\sqrt{2}}%
\begin{pmatrix}
1 & \mp e^{-i\theta } \\
\pm e^{i\theta } & 1%
\end{pmatrix}%
,
\end{equation}%
where $\theta $ is a real parameter. Let us define the eigenstates
of $H(t)$ at time $T$ as $|\phi _{n}(T)\rangle =V_{\theta \pm
}|n\rangle $, $n\in \{0,1\}$. We can then write $U_{A_{n}}(T)=e^{i\alpha
_{n}}V_{\theta \pm }|n\rangle \langle n|$, where
$\alpha
_{n}$ are geometric phases, and {we can show} that $e^{i\alpha _{0}}=e^{i\alpha _{1}}$ \cite%
{Ore08}. Therefore, up to a global phase, Eq.~\eqref{finalU} yields $%
U(T)=V_{\theta \pm }$. Using the last result and the identity $\pm
(\cos{\theta}X+\sin{\theta}Y)=H_{\theta\pm}$, one can verify the
following constructions: an adiabatic interpolation along
the path $-Z\otimes \widetilde{%
G}\rightarrow -X\otimes \widetilde{G}$ yields the operation
$V_{0+}=RZ$ where $R$ is the Hadamard gate; an adiabatic
interpolation
along the path $-Z\otimes \widetilde{G}\rightarrow -(\frac{1}{\sqrt{2}}X+%
\frac{1}{\sqrt{2}}Y)\otimes \widetilde{G}\rightarrow Z\otimes
\widetilde{G}$ yields $V^{\dagger}_{\pi/4-}V_{\pi/4+}=V_{\pi/4+}^2$,
which up to an overall phase is equal to $XS$ where $S$ is the Phase
gate. $RZ$ and $XS$ can generate all
single-qubit operations in the Clifford group. Note that
single-qubit gates outside the Clifford group can also be
implemented in a similar manner \cite{Ore08}. {We point} out that
the change of the Hamiltonian along the edges of the paths {must} be
sufficiently slow so that the adiabatic approximation \cite{HJ02} is
satisfied within the desired precision; we return to the validity of
the adiabatic approximation below.

\textit{Completing the gate set}.---To complete the set of gates
needed for encoded Clifford operations, we only have to show how to
implement a transversal CNOT gate with the first qubit being the
target. {At the moments between}
the basic operations building up an encoded Clifford gate, we can always find $%
\widetilde{G}$ which acts trivially on the control qubit. Then the
CNOT gate can be applied by first applying the inverse of the Phase
gate on the control, as well as the transformation $-Z\otimes
\widetilde{G}\rightarrow -Y\otimes \widetilde{G}$ on the target, followed by the transformation $%
-I^{c}\otimes Y\otimes \widetilde{G}\rightarrow -Z^{c}\otimes
Z\otimes \widetilde{G}$ where the superscript $c$ denotes the
control.

Encoded operations outside of the Clifford group require the above
transformations plus the ability to measure a particular encoded
Clifford operator \cite{Got97}. The latter involves applying the
operator conditioned on the qubits in a \textquotedblleft
cat\textquotedblright\ state $(|0...0\rangle +|1...1\rangle
)/\sqrt{2}$. If we were implementing solely the Clifford operator,
at any stage between the elementary gates we would use a stabilizer
or gauge group element of the form $\widehat{G}=G_{1}\otimes
G_{\bar{1}}$, where $G_{1}$ is a tensor product of Pauli matrices
acting on the first qubits from the blocks, and $G_{\bar{1}}$ is an
operator on the rest of the qubits. Applying a basic transversal
operation $O$ conditioned on the first qubit in a cat
state transforms this operator as $I^{c}\otimes G_{1}\otimes G_{\bar{1}%
}\rightarrow |0\rangle \langle 0|^{c}\otimes G_{1}\otimes G_{\bar{1}%
}+|1\rangle \langle 1|^{c}\otimes OG_{1}O^{\dagger }\otimes G_{\bar{1}}$,
where the superscript $c$ denotes the control qubit from the cat state. The
corresponding gate can be implemented via the Hamiltonian $\widehat{\widehat{%
H}}_{O}(t)=-|0\rangle \langle 0|^{c}\otimes G_{1}\otimes G_{\bar{1}}-\alpha
(t)|1\rangle \langle 1|^{c}\otimes H_{O}(t)\otimes G_{\bar{1}}$, where $%
H_{O}(t)\otimes G_{\bar{1}}$ is the Hamiltonian that we would use for the
implementation of the operation $O$, and $\alpha (t)$ is chosen so that the
operator $\alpha (t)|1\rangle \langle 1|^{c}\otimes H_{O}(t)\otimes G_{\bar{1%
}}$ has the same instantaneous spectrum as $|0\rangle \langle 0|^{c}\otimes
G_{1}\otimes G_{\bar{1}}$. (Any possible relative geometric phase between $%
|0\rangle ^{c}$ and $|1\rangle ^{c}$ can be corrected by a suitable
gate on the control qubit.) Starting from
$\widehat{\widehat{H}}_{O}(T)$, we can implement another conditional
transversal operation in a similar fashion, etc.

\textit{Fault tolerance of the scheme}.---{So far we have shown how
to} generate any transversal operation on an encoded state
holonomically, assuming that the state is non-erroneous. But what if
an error occurs on one of the qubits? At any time $t$, we can
distinguish two types of errors:\
those that result in transitions between the ground and excited spaces of $%
H(t)$, and those that result in transformations inside the
eigenspaces. Due to the discretization of errors in quantum error
correction (QEC), it suffices to prove correctability for each type
separately. The key property of our construction is that the
geometric transformation is the same in each of the eigenspaces, and
it is transversal. Because of this, if we are applying a unitary on
the first qubit, an error on that qubit will remain localized
regardless of whether it causes an excitation or not. If the error
occurs on one of the other qubits, at the end of the transformation
the result would be the desired single-qubit unitary gate plus the
error on the other qubit, which is correctable. It is remarkable
that even though the Hamiltonian couples qubits within the same
block, single-qubit errors do not propagate. This is because the
coupling between the qubits amounts to a change in the relative
phase between the ground and excited spaces, but the latter is
irrelevant since it is either equivalent to a gauge transformation,
or when we apply a correcting operation we project on one of the
eigenspaces. In the case of the CNOT gate, an error can propagate
between the control and the target qubits, but it never results in
two errors within the same codeword.

In addition to transversal operations, a complete fault-tolerant
scheme requires the ability to prepare, verify and use special
ancillary states, e.g., Shor's cat state \cite{FT}. Since the cat
state is known and its construction is non-fault-tolerant, this can
always be done using our holonomic approach by treating each
initially prepared qubit as a simple code (with $\widetilde{G}$
being trivial), and updating the stabilizer of the code via the
applied geometric transformation as the operation progresses. The
only difference is in the measurement of the parity of the state,
which requires the ability to apply successively CNOT operations
from two different qubits in the cat state to one and the same
ancillary qubit initially prepared in the state $|0\rangle $. After
this operation, the ancilla qubit is measured in the $\{|0\rangle $,
$|1\rangle\} $ basis so the relative phase between these two states
is irrelevant. We can regard the qubit in state $|0\rangle $ as a
simple code with stabilizer $\langle Z\rangle $, and we can apply
the first CNOT as described before. Even though after this operation
the state of the target qubit is unknown, the second CNOT can be
applied via the same interaction, since the transformation in each
eigenspace of the Hamiltonian is the same.

\textit{Discussion}.---Since the method we presented conforms
completely to a {given} fault-tolerant scheme, it {does} not affect
the error threshold per operation for that scheme. Some of its
features, however, affect the threshold for \emph{environment}
noise. First, observe that when applying the Hamiltonian
\eqref{Ham1}, we cannot at the same time apply operations on the
other qubits on which the factor $\widetilde{G}$ acts non-trivially.
This could be a problem for non-concatenation schemes, but with
concatenated codes it only affects the lowest level of
concatenation---some operations at the lowest level that would
otherwise be implemented simultaneously might have to be implemented
serially. {This has the effect of slowing down the
circuit by a small constant factor.} For example, {it turns out} that for the $9$%
-qubit Bacon-Shor (BS) subsystem code \cite{Bac06} this slowdown is
by a factor of $1.5$ \cite{Ore08}.

A more significant slowdown results from the fact that the evolution is
adiabatic (however, as argued in Ref. \cite{ALZ06}, an adiabatic
implementation may be unavoidable in the case of Markovian noise, since fast
gates are incompatible with the Markovian limit). In order to obtain a rough
estimate of the slowdown due specifically to the adiabatic requirement, we
compare the time $T_{h}$ needed for the implementation of a holonomic gate
with precision $1-\delta $ to the time $T_{d}$ needed for a dynamical
realization of the same gate with the same strength of the Hamiltonian. We
consider a realization of the $X$ gate via the Hamiltonian ${H}%
(t)=V_{X}(y(t))ZV_{X}^{\dagger }(y(t))$, $V_{X}(y(t))=\exp (iy(t)\frac{\pi }{%
2T_{h}}X)$, where $y(0)=0$, $y(T_{h})=T_{h}$. The energy gap of
this Hamiltonian is constant. The optimal dynamical implementation
of the same gate is via the Hamiltonian $-X$ for time
$T_{d}=\frac{\pi }{2}$. It is known that if $H(t)$ is smooth (but
non-analytic) and its derivatives vanish at $t=0$ and $t=T_{h}$, the
adiabatic error decreases
super-polynomially with $T_{h}$ \cite{HJ02}. To achieve this, we choose $%
y(t)=\frac{1}{a}\int_{0}^{t}dt^{\prime }e^{-1/\sin (\pi t^{\prime }/T_{h})}$%
, $a=y(T_{h})$. For this interpolation, when $T_{h}/T_{d}\approx 17$, the
error $\delta $ is already $\sim 10^{-6}$, which is below the threshold
values obtained, e.g., for the BS codes ($\sim 10^{-4}$) \cite{AC07}. This
slowdown would decrease the allowed rate of environment decoherence by a
similar factor. But dynamical gates are not perfect in practice, and
the holonomic approach may be advantageous if it leads to {higher}
operational precision.

In Ref.~\cite{Ore08} we show that for the BS code our scheme can be
implemented with at most $3$-local Hamiltonians. This is optimal for the
construction we presented, since there are no codes correcting
arbitrary single-qubit errors with stabilizer or gauge-group elements of weight smaller than $2$, covering
all qubits. An interesting {open} question is whether it is possible
to modify our construction so that it uses $2$-local interactions, perhaps
using recent perturbative \textquotedblleft gadget\textquotedblright\ ideas
\cite{Jordan08}.

{Our abstract scheme proves that the holonomic quantum computing
approach is scalable under a reasonable noise model. It is meant as
a proof of principle, and will undoubtedly require modifications if
applied to actual physical systems}. Given that simple QECCs and two-qubit geometric transformations have been realized using NMR \cite%
{NMR} and ion-trap \cite{Tra} techniques, these systems seem particularly
suitable for hybrid HQC-QEC implementations.

{Acknowledgements.---Research supported by NSF under grants
CCF-0524822 (to OO), CCF-0448658 (to TAB), and CCF-0523675 (to
DAL).} The authors thank Paolo Zanardi for illuminating discussions.


\begin{thebibliography}{99}
\bibitem{HQC} P. Zanardi and M. Rasetti, Phys. Lett. A \textbf{264}, 94
(1999); J. Pachos, P. Zanardi, and M. Rasetti, Phys. Rev. A \textbf{61},
010305(R) (1999).

\bibitem{Berry84} M. Berry, Proc. R. Soc. Lond. A \textbf{392}, 45 (1984).

\bibitem{QCmodels} Circuit model:\ D. Deutsch, Proc. Roy. Soc. London Ser. A
425, \textbf{73} (1989); One-way model:\ R. Raussendorf and H.J. Briegel,
Phys. Rev. Lett. \textbf{86}, 5188 (2001); Adiabatic QC model:\ E. Farhi
\textit{et al}. eprint quant-ph/0001106.

\bibitem{WZ84} F. Wilczek and A. Zee, Phys. Rev. Lett. \textbf{52}, 2111
(1984).

\bibitem{HQCrob} A. Carollo \textit{et al.}, Phys. Rev. Lett. \textbf{90},
160402 (2003); G. De Chiara and G. M. Palma, Phys. Rev. Lett. \textbf{91},
090404 (2003);
%P. Solinas, P. Zanardi, and N. Zangh, Phys. Rev. A \textbf{70},
%042316 (2004);
I. Fuentes-Guridi, F. Girelli, and E. Livine, Phys. Rev. Lett
\textbf{94}, 020503 (2005); S.-L. Zhu and P. Zanardi, Phys. Rev. A
\textbf{72}, 020301(R) (2005); G. Florio \textit{et al.}, Phys. Rev.
A \textbf{73}, 022327 (2006).

\bibitem{SarandyLidar:06} M.S. Sarandy and D.A. Lidar, Phys. Rev. A \textbf{%
73}, 062101 (2006).

\bibitem{Got97} D. Gottesman, Phys. Rev. A \textbf{57}, 127 (1998).

\bibitem{FT} E. Knill, R. Laflamme and W. Zurek, Science \textbf{279}, 342
(1998); A.M. Steane, Phys. Rev. A \textbf{68}, 042322 (2003); P. Aliferis,
D. Gottesman, J. Preskill, Quant. Inf. Comput. \textbf{6}, 97 (2006).

\bibitem{AL06} P. Aliferis and D.W. Leung, Phys. Rev. A \textbf{73}, 032308 (2006).

\bibitem{QEC} P. W. Shor, Phys. Rev. A \textbf{52}, R2493 (1995); A. M. Steane,
Phys. Rev. Lett. \textbf{77}, 793 (1996).

\bibitem{WZL05} L.-A. Wu, P. Zanardi, D. A. Lidar, Phys. Rev. Lett. \textbf{%
95}, 130501 (2005).

\bibitem{DFS} P. Zanardi and M. Rasetti, Phys. Rev. Lett. \textbf{79}, 3306
(1997); D. A. Lidar, I. L. Chuang, and K. B. Whaley, Phys. Rev. Lett.
\textbf{81}, 2594 (1998).

\bibitem{Stab} D. Gottesman, Phys. Rev. A \textbf{54}, 1862 (1996); A. R.
Calderbank \textit{et al.}, Phys. Rev. Lett. \textbf{78}, 405 (1997).

\bibitem{Bac06} D. Bacon, Phys. Rev. A \textbf{73}, 012340 (2006).

\bibitem{OQEC} D. W. Kribs, R. Laflamme, D. Poulin, Phys. Rev. Lett. \textbf{%
94}, 180501 (2005); D. Poulin, Phys. Rev. Lett. \textbf{95}, 230504 (2005).

\bibitem{Ore08} O. Oreshkov, T. Brun, and D.A. Lidar, in preparation.

%\bibitem{KAS06} D. Kult, J. {\AA }berg, and E. Sj\"{o}qvist, Phys. Rev. A
%\textbf{74}, 022106 (2006).

\bibitem{HJ02} G. A. Hagedorn and A. Joye, J. Math. Anal. and Appl. \textbf{%
267}, 235 (2002).

\bibitem{ALZ06} R. Alicki, D. A. Lidar, P. Zanardi, Phys. Rev. A \textbf{73}%
, 052311 (2006).

\bibitem{AC07} P. Aliferis and A. W. Cross, Phys. Rev. Lett. \textbf{98},
220502 (2007).

\bibitem{Jordan08} S. Jordan and E. Farhi, Phys. Rev. A \textbf{77}, 062329 (2008).

\bibitem{NMR} D. G. Cory {\textit{e}t\textit{\ al}.}, Phys. Rev. Lett.
\textbf{81}, 2152 (1998); J. A. Jones {\textit{e}t \textit{al}.}, Nature
\textbf{403}, 869 (2000).

\bibitem{Tra} D. Leibfried {\textit{e}t \textit{al}.}, Nature \textbf{422},
412 (2003); J. Chiaverini {\textit{e}t \textit{al}.}, Nature \textbf{432},
602 (2004).
\end{thebibliography}
\end{document}